\documentclass[aps,prd,twocolumn,showpacs,amsmath,amssymb]{revtex4-1}
\usepackage{amsmath}
\usepackage{graphicx}
\usepackage{subfigure}
\usepackage{epsfig}
\usepackage{epstopdf}
\usepackage{color}
\usepackage{multirow}
\usepackage{setspace}
\usepackage{overpic}
\usepackage{amssymb}
\usepackage[driverfallback=dvipdfmx, bookmarksnumbered, pdfstartview=FitH,colorlinks,urlcolor=blue, citecolor=blue,linkcolor=blue] {hyperref}
\usepackage{bm}
\usepackage{rotating}
\usepackage[utf8]{inputenc}

\usepackage{verbatim}

\newcommand{\jpsi}{J/\psi}

\newcommand{\LLb}{\Lambda\bar{\Lambda}}

\newcommand{\Lamb}{\bar{\Lambda}}

\newcommand{\III}{\uppercase\expandafter{\romannumeral3}}
\newcommand{\II}{\uppercase\expandafter{\romannumeral2}}

\newcommand{\btable}{\begin{table}}
\newcommand{\etable}{\end{table}}
\newcommand{\btu}{\begin{tabular}}
\newcommand{\etu}{\end{tabular}}
\newcommand{\bfigure}{\begin{figure}}
\newcommand{\efigure}{\end{figure}}
\newcommand{\bitem}{\begin{itemize}}
\newcommand{\eitem}{\end{itemize}}

\newcommand{\gpp}{\gamma\pi^+\pi^-}
\newcommand{\epp}{\eta\pi^+\pi^-}
\newcommand{\LLE}{\Lambda\bar{\Lambda}\eta'}

\begin{document}


\title{\boldmath Observation of $\psi(3686) \rightarrow \Lambda \bar{\Lambda} \eta^{\prime}$ decay}

\author{M.~Ablikim$^{1}$, M.~N.~Achasov$^{5,b}$, P.~Adlarson$^{75}$, X.~C.~Ai$^{81}$, R.~Aliberti$^{36}$, A.~Amoroso$^{74A,74C}$, M.~R.~An$^{40}$, Q.~An$^{71,58}$, Y.~Bai$^{57}$, O.~Bakina$^{37}$
% ...
\\
\vspace{0.2cm}
(BESIII Collaboration)\\
\vspace{0.2cm} {\it
$^{1}$ Institute of High Energy Physics, Beijing 100049, People's Republic of China\\
% ...
$^{5}$ Budker Institute of Nuclear Physics SB RAS (BINP), Novosibirsk 630090, Russia\\
% ...
$^{36}$ Johannes Gutenberg University of Mainz, Johann-Joachim-Becher-Weg 45, D-55099 Mainz, Germany\\
$^{37}$ Joint Institute for Nuclear Research, 141980 Dubna, Moscow region, Russia\\
% ...
$^{40}$ Liaoning Normal University, Dalian 116029, People's Republic of China\\
% ...
$^{57}$ Southeast University, Nanjing 211100, People's Republic of China\\
$^{58}$ State Key Laboratory of Particle Detection and Electronics, Beijing 100049, Hefei 230026, People's Republic of China\\
% ...
$^{71}$ University of Science and Technology of China, Hefei 230026, People's Republic of China\\
% ...
$^{74}$ University of Turin and INFN, (A)University of Turin, I-10125, Turin, Italy; (B)University of Eastern Piedmont, I-15121, Alessandria, Italy; (C)INFN, I-10125, Turin, Italy\\
$^{75}$ Uppsala University, Box 516, SE-75120 Uppsala, Sweden\\
% ...
$^{81}$ Zhengzhou University, Zhengzhou 450001, People's Republic of China\\
\vspace{0.2cm}
% ...
$^{b}$ Also at the Novosibirsk State University, Novosibirsk, 630090, Russia\\
% ...
}
}

\begin{abstract}
 Using a sample of $(27.12\pm 0.14) \times 10^{8}$ 
 $\psi(3686)$ events collected with the BESIII detector, the decay $\psi(3686) \to \Lambda \bar{\Lambda} \eta'$ with $\eta'$ subsequently decaying into $\gamma \pi^+ \pi^-$ and $\eta \pi^+ \pi^-$ is observed for the first time. The branching fraction of $\psi(3686) \to \Lambda \bar{\Lambda}\eta'$ is measured 
to be $(7.34\pm0.94(stat.)\pm0.43(sys.))\times10^{-6}$.
No resonant structures are  evident in the $\Lambda\eta'$, $\Lamb\eta'$ and  $\Lambda\Lamb$ mass spectra.
\end{abstract}

\maketitle

\oddsidemargin  -0.2cm
\evensidemargin -0.2cm


\section{Introduction}

Quantum Chromodynamics (QCD), the theory describing
the strong interaction, has been tested thoroughly at high
energy. However, in the medium-energy region, theoretical
calculations based on first principles are still unreliable since
the non-perturbative contribution is significant and  models must be employed.
Charmonium resonances, as bound states of a charm and an anti-charm quark, are governed by  long-range 
interactions and exist on the boundary between the perturbative
and non-perturbative regimes in QCD~\cite{Kwong1987mj,Eichten2007qx}.
Therefore, the study of charmonium decays  
can provide valuable inputs for gaining a better understanding of the structure of these states and for improving our knowledge of QCD.
The BESIII experiment has collected large data samples of  $J/\psi$ and $\psi(3686)$ events, 
making possible the study of decay channels of these states, many with complicated intermediate structures.

The decays of $J/\psi$ and $\psi(3686)$ mesons into baryon pairs have been understood in terms of $c\bar{c} $ annihilations into three gluons or a virtual photon~\cite{HP}.
However, three-body decays, for example $J/\psi(\psi(3686))\to\LLb P$, where $``P"$ represents a pseudoscalar meson such as $\pi^{0},\eta$ or $\eta'$, warrant further study because of the potentially important contribution of intermediate states.
The $\Lambda(1670)$ has been found to make a significant contribution to the decay $\psi(3686)\to\LLb\eta$~\cite{2022ws,2013ablikim}.
However, the decay of  $J/\psi\rightarrow \Lambda\bar{\Lambda}\eta^\prime$ is hard to observe experimentally because of the limited phase space, 
while the study of $\psi(3686)\rightarrow  \Lambda\bar{\Lambda}\eta^\prime$  
is feasible but  has not been attempted yet.

Since the excitation spectra of most hyperons are still not well understood~\cite{ex,Hex-2020}, it is important to search for excited hyperons which have not yet been observed.
The decay $\psi(3686)\rightarrow 
\Lambda\bar{\Lambda}\eta^\prime$ provides an opportunity to search for potential $\Lambda$ excitations. A sample of  $(27.12\pm 0.14) \times 10^{8}$ $\psi(3686)$ events~\cite{psinum} produced in $e^+e^-$ 
annihilations~\cite{2022pdg} has been collected with the BESIII detector, which allows for the experimental study of the decay   $\psi(3686)\rightarrow \Lambda\bar{\Lambda}\eta^\prime$.

\section{Detector and Monte Carlo Simulation}
The BESIII detector~\cite{Ablikim:2009aa} records symmetric $e^+e^-$ collisions 
provided by the BEPCII storage ring~\cite{Yu:IPAC2016-TUYA01}
in the center-of-mass energy range from 2.0 to 4.95~GeV, with a peak luminosity of $1 \times 10^{33}\;\text{cm}^{-2}\text{s}^{-1}$ 
achieved at $\sqrt{s} = 3.77\;\text{GeV}$.  BESIII has collected large data samples in this energy region~\cite{Ablikim:2019hff} \cite{EventFilter}.
The cylindrical core of 
the BESIII detector consists of a helium-based main drift chamber (MDC), a plastic 
scintillator time-of-flight (TOF) system, and a CsI(Tl) electromagnetic calorimeter (EMC), 
which are all enclosed in a super conducting solenoidal magnet providing a 1.0~T magnetic field. The solenoid is supported by an octagonal flux-return yoke with 
resistive plate counter modules interleaved with steel for muon identification. The acceptance for charged particles and photons is 93\% of the full solid angle, and the charged-particle 
momentum resolution at 1~GeV/c is 0.5\%. The photon energy resolution is 2.5\% (5\%) at 1.0 
GeV in the barrel (end-cap) region.
The time resolution in
the TOF barrel region is 68 ps, while that in the end-cap region is 110 ps. The end-cap TOF system was upgraded in 2015 using multigap resistive plate chamber technology, 
providing a time resolution of 60~ps,
which benefits $\sim$84\% of the data used in this analysis~\cite{tof2017lix,tof2017caop,tof2020caop} .

 \par
 Simulated data samples produced with a \textsc{geant4}-based  Monte Carlo (MC) 
 package~\cite{2003Agostinelli}, which includes the geometric 
 description of the BESIII detector and the detector response, are used to determine 
 detection efficiencies and to estimate backgrounds. The simulation models the beam-energy 
 spread and initial-state radiation (ISR)  in  $e^{+}e^{-}$ annihilations with the 
 generator \textsc{kkmc}~\cite{kkmc2000,kkmc2001}. 
An inclusive MC sample of  2.7 billion  $\psi(3686)$ events is used to investigate potential background. The inclusive MC sample includes the production 
 of the  $\psi(3686)$ resonance, the ISR production of the $J/\psi$, and
the continuum processes incorporated in {\sc
kkmc}. The known particle decays are modelled with 
 \textsc{evtgen}~\cite{2001Lange,2008ping}  using branching fractions taken from the 
 Particle Data Group~\cite{2022pdg}, while the remaining unknown decays are estimated with \textsc{lundcharm}~\cite{2000chen,lund2014}.

To  optimize the selection criteria and determine the detection efficiency, the signal MC sample of $2.6\times10^{6}$ $\psi(3686)\to\Lambda\bar\Lambda\eta'$ events is generated  with uniform phase space (PHSP), where the process
 $\eta' \to \gamma \pi^+ \pi^-$ is described according to
theoretical models~\cite{mode1-gammapipi,mode2-gammapipi} that have been validated in previous measurements~\cite{bes3-gammapipi}, and the process $\eta' \to \eta \pi^+ \pi^-$ is simulated
according to the distributions measured in
Ref.~\cite{mode1-etaptoetapipi}.
The data sample taken at the center of mass (CM) energy of 3.773 GeV, corresponding to an integrated luminosity of  $(2916.94\pm0.18\pm29.17)$ pb$^{-1}$
~\cite{ctn,ctn2}, is  used  to  estimate the background events directly from $e^+e^-$ annihilations.
  
\section{event selection}
In this analysis, the decay $\psi(3686)\to\Lambda\bar\Lambda\eta'$ is selected by reconstructing one $\Lambda\bar\Lambda$ pair and one $\eta'$ meson, where the $\Lambda(\bar{\Lambda})$ candidate is reconstructed by the $p\pi^-(\bar{p}\pi^+)$ decay, and the $\eta'$ candidate is reconstructed by its two dominant decay modes, $ \eta' \to \gamma \pi^+ \pi^-$ ({\bf Mode I}) and  $\eta' \to \eta \pi^+   \pi^-$ ({\bf Mode II}).
Thus, the reconstructed final states for $\psi(3686) \to \Lambda \bar{\Lambda} \eta'$ are $p\bar{p}\pi^+\pi^-\pi^+\pi^-\gamma$  and $p\bar{p}\pi^+\pi^-\pi^+\pi^-\gamma\gamma$.

The number of charged tracks is required to be more than five. Each track must satisfy $|\cos \theta| < 0.93$, where $ \theta$ is the polar angle with respect to the MDC symmetry axis.  
Each photon candidate is required to have a deposited energy in the EMC more than 25 MeV in the barrel region ($|\cos \theta|< 0.80$) and more than 50 MeV in the end-cap region ($0.86 < |\cos \theta| <0.92$). To exclude showers arising from charged tracks, the angle between the EMC shower and the position of the closest charged track at the EMC
 must be greater than 10 degrees as measured from the IP. To suppress electronic noise and showers unrelated to the event, the difference between the EMC time and the event-start time is required to be within [0, 700] ns. At least one photon candidate is required for  {\bf Mode I}, and at least two photon candidates for {\bf Mode II}.

For each charged track, the information from both the TOF and d$E/$d$x$  are combined to form a particle identification (PID) probability for the $\pi$, $K$, and $p$ hypotheses (Prob($i$), $i$ = $\pi$, $K$, $p$). A charged track is identified as a pion or proton if its probability is greater than that for any other assignments.

The $\Lambda$ ($\bar{\Lambda}$) candidate is reconstructed from pairs of (anti-)proton and oppositely charged pion tracks fulfilling a secondary vertex fit. Events with at least one $\Lambda$ and one $\bar{\Lambda}$ 
candidate are selected.  When looping over all combinations of $\Lambda$ and $\bar{\Lambda}$ candidates, the one with the minimal value of $(M(p\pi^{-})-M(\Lambda))^2+(M(\bar{p}\pi^{+})-M(\bar{\Lambda}))^{2}$ is chosen,
 where $M(\Lambda)$ ($M(\bar{\Lambda})$) is the known mass of the $\Lambda$ ($\bar{\Lambda}$) baryon~\cite{2022pdg}. The remaining charged pions not associated to the $\Lambda(\bar{\Lambda})$ candidates are considered as coming from $\eta^\prime$ decays, and are 
required to originate from a region of 10~cm around the interaction point along the beam direction and 1~cm in the plane perpendicular to the beam.
The pairs of opposite charged pions are requested to pass a primary vertex fit under the $\pi^{+}\pi^{-}\Lambda\bar{\Lambda}$ hypothesis, 
and that combination with the smallest 
$\chi^2$ is retained.
  
To improve the momentum resolution and reduce the background, a kinematic fit is applied to the event candidates. For {\bf Mode I}, the conservation of the initial-state energy and momentum is required (four constraints); for {\bf Mode II}, in addition to four-momentum conservation, the invariant mass of the photon pair is constrained to the known $\eta$ mass. 
For events with more than one or two photons, respectively for {\bf Mode I} and {\bf Mode II}, all the combinations are fitted and that with the best fit quality is selected.
We further require  the fit quality to satisfy $\chi^{2}_{4C}<30$ for {\bf Mode I} and $\chi^{2}_{5C}<40$  for {\bf Mode II},
a selection that is optimized by maximizing the figure of merit $S/\sqrt{S + B}$, where S is the number of
signal MC events obtained from MC simulation 
and $S+B$ is the number of signal and background events obtained from real data.

\begin{figure}[htp]
 \centering
  \begin{overpic}[width=0.5\textwidth]{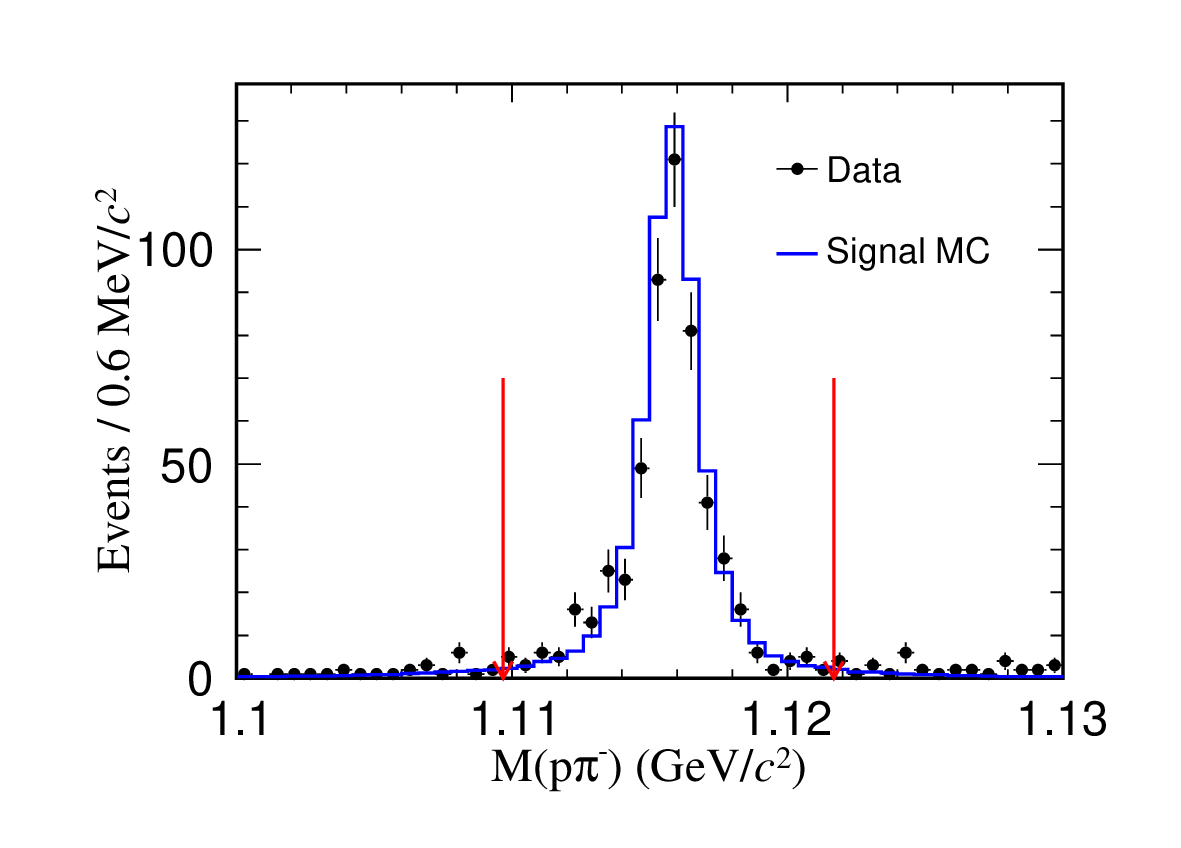}
  \put(30,55){$(a)$}
  \end{overpic}
  \begin{overpic}[width=0.5\textwidth]{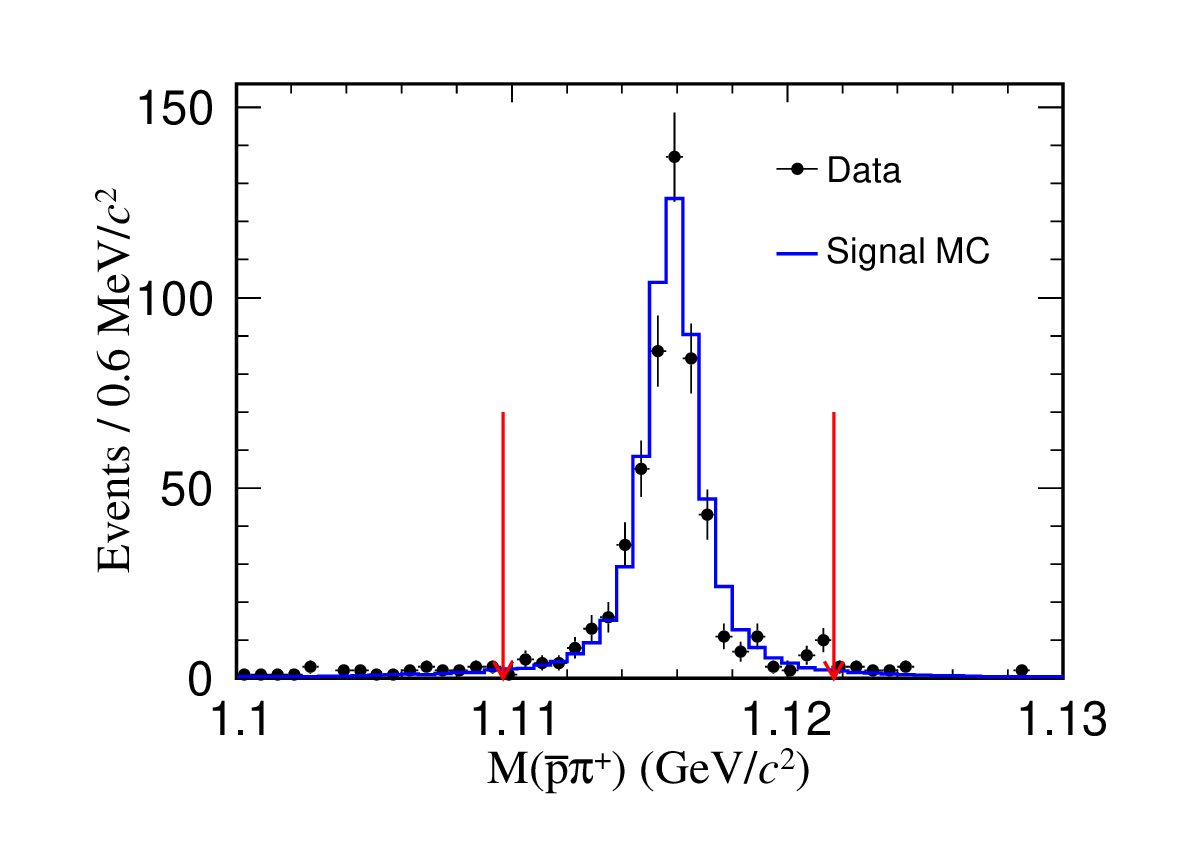}
  \put(30,55){$(b)$}
  \end{overpic}
    \caption{Distributions of $M_{p\pi^{-}}$ (a) and $M_{\bar{p}\pi^{+}}$ (b) for {\bf Mode I}, where the arrows indicate the region within which the  $\Lambda/\bar{\Lambda}$ signal is selected.}
 \label{fig:mlam1}
  \end{figure}

\begin{figure}[htpb]
 \begin{overpic}[width=0.5\textwidth]{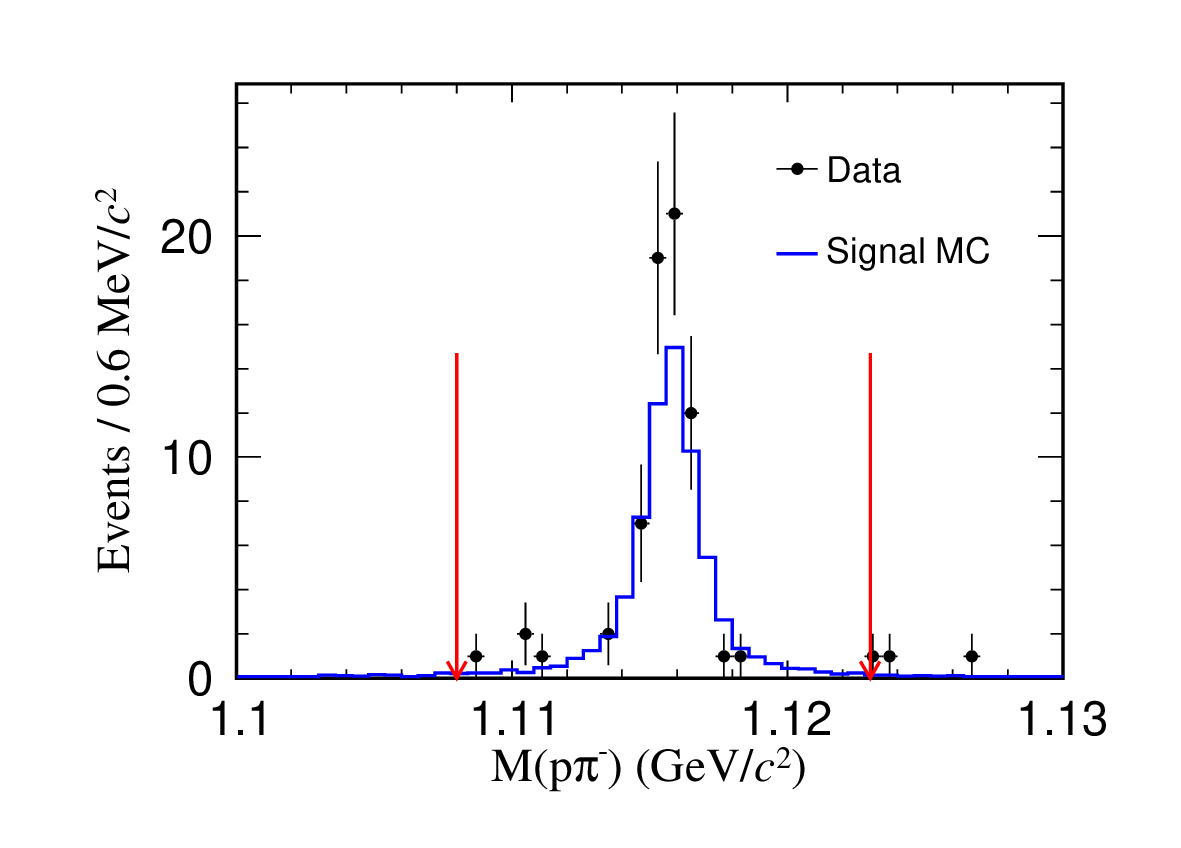}
  \put(30,55){$(a)$}
  \end{overpic}
  
  \begin{overpic}[width=0.5\textwidth]{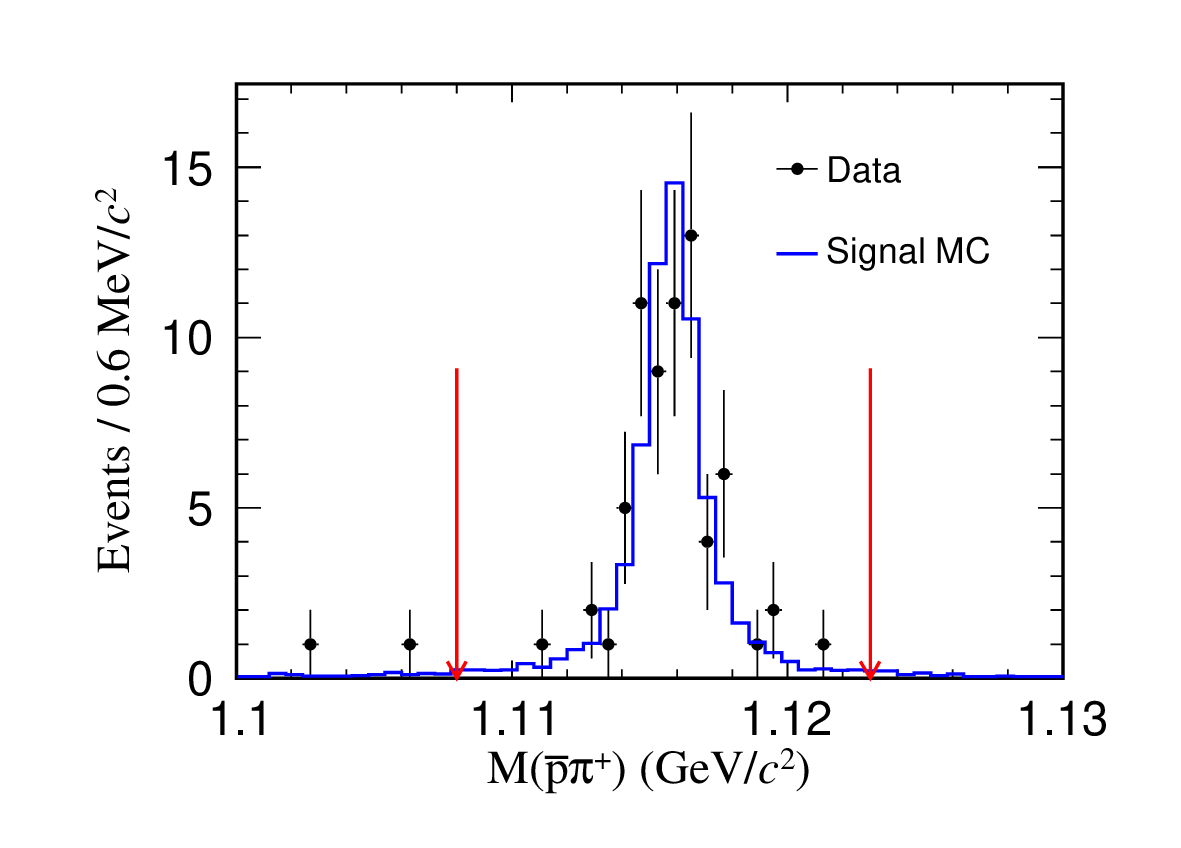}
  \put(30,55){$(b)$}
  \end{overpic}
 \caption{Distributions of $M_{p\pi^{-}}$ (a) and $M_{\bar{p}\pi^{+}}$ (b) for {\bf Mode II}, where the arrows indicate the region within which the  $\Lambda/\bar{\Lambda}$ signal is selected.}
 \label{fig:mlam2}

\end{figure}

Events with candidate $\eta'$ decays are kept for further analysis,  by selecting the invariant-mass signal regions 
$0.94<M(\gpp)<0.97 $ GeV/$c^{2}$ and 
$|M(\epp)-M(\eta')|<10$ MeV/$c^{2}$ for {\bf Mode I} and {\bf Mode II}, respectively. The invariant-mass distributions $M(p\bar \pi)$ and $M(\bar p\pi^-)$ for the two modes
are shown in  Fig.~\ref{fig:mlam1} and Fig.~\ref{fig:mlam2},  where the $\Lambda$ and $\bar{\Lambda}$ peaks are clearly visible.

\begin{figure}[htp]
 \centering
  \begin{overpic}[width=0.5\textwidth]{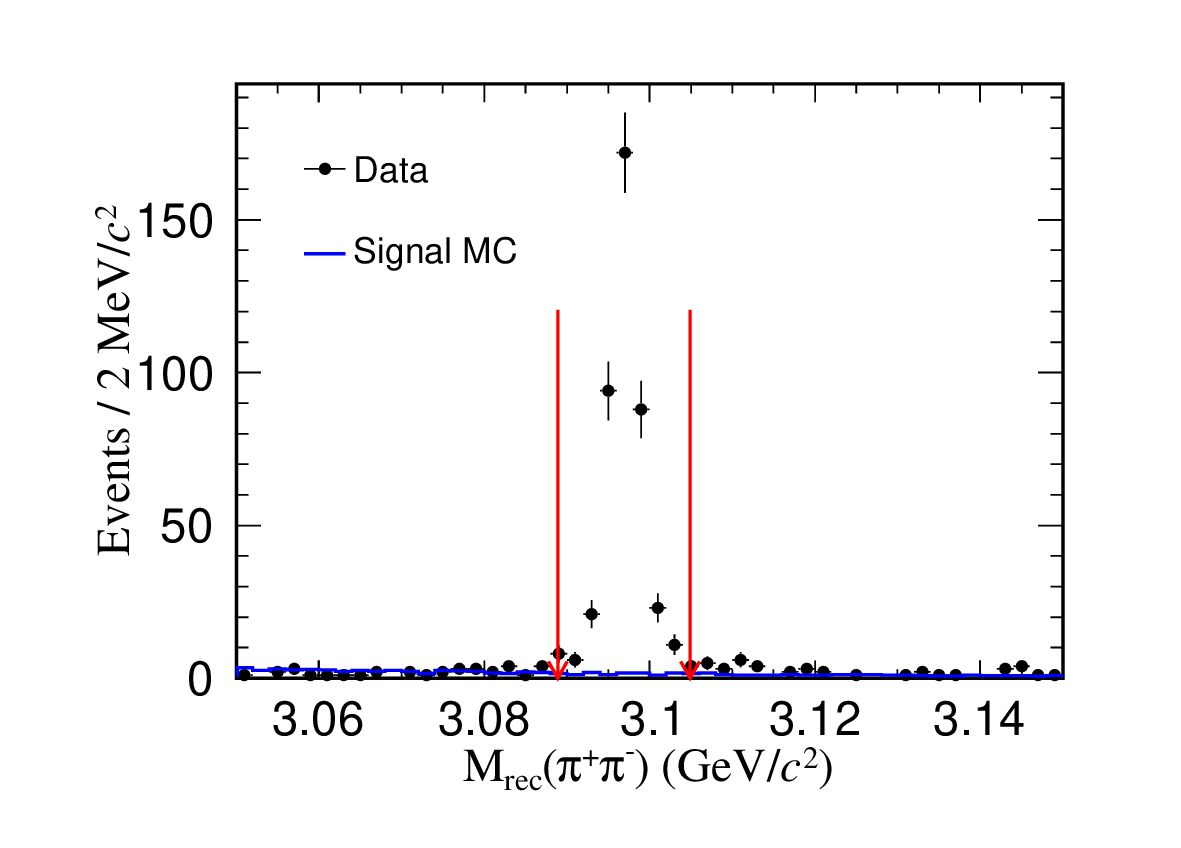}
  \put(70,55){$(a)$}
  \end{overpic}
 
  \begin{overpic}[width=0.5\textwidth]{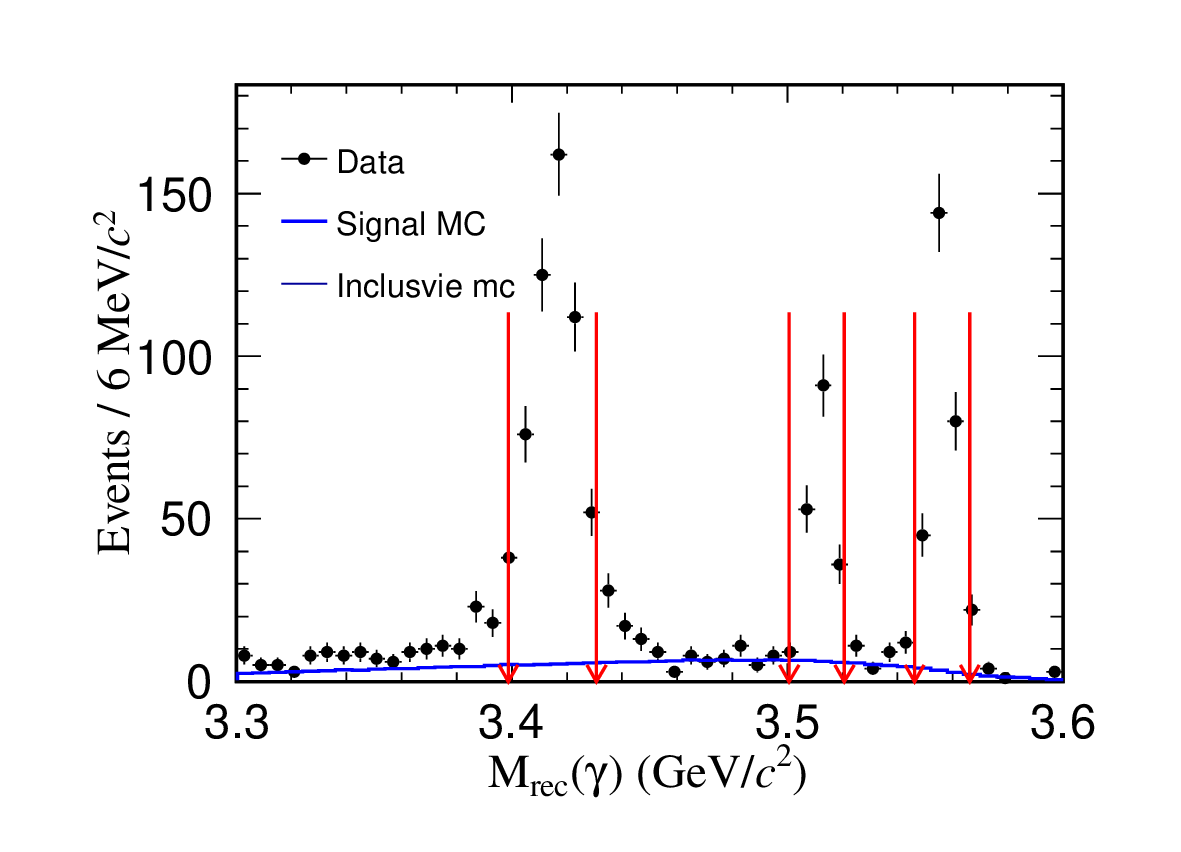}
  \put(70,55){$(b)$}
  \end{overpic}
  \caption{Distributions of  $M_{\rm rec}(\pi^{+}\pi^{-})$ (a) and $M_{\rm rec}(\gamma)$ (b) for Mode I, where the solid arrows indicate the mass window within which (a) $J/\psi$ and (b) $\chi_{c0/c1/c2}$ decays are vetoed.   The signal MC is normalised to the measured branching fraction.
  }
  \label{fig:gpp}
\end{figure}

For  {\bf Mode I}, the $\Lambda(\Lamb)$ candidates are selected by requiring 
$|M(p\pi^{-}/\bar{p}\pi^{+})-M(\Lambda/\Lamb)|< 6$ MeV/$c^{2}$. This signal region is indicated by arrows in  Fig.~\ref{fig:mlam1}.
In order to remove  background events from $\psi(3686)\to\Sigma^{0}\Lamb\pi^{+}\pi^{-} + c.c.$, we require $|M(\gamma\Lambda/\gamma\bar{\Lambda})-M(\Sigma^{0}/\bar{\Sigma}^{0})| >10$ MeV/$c^{2}$. 
The distribution of the recoil mass against the $\pi^+\pi^-$ system, $M_{\rm rec}(\pi^+\pi^-)$, is shown in Fig.~\ref{fig:gpp}(a), where the $J/\psi$ peak is visible due to the background events from the $\psi(3686)\rightarrow\pi^+\pi^- J/\psi$ process. To veto this background, the recoil mass is required to satisfy $|M_{\rm rec}(\pi^{+}\pi^{-}) -M(\jpsi)| >8$ MeV/$c^{2}$.
The recoil mass against the $\gamma$, $M_{\rm rec}(\gamma)$, is shown in Fig.~\ref{fig:gpp}(b), where the three peaks due to the $\chi_{c0}$, $\chi_{c1}$ and $\chi_{c2}$ mesons are visible. To suppress these background events,  events in the mass regions $|M_{\rm rec}(\gamma)-M(\chi_{c0})|< 16$ MeV/$c^{2}$, $|M_{\rm rec}(\gamma)-M(\chi_{c1})|< 10$ MeV/$c^{2}$ and $|M_{\rm rec}(\gamma)-M(\chi_{c2})|< 10$ MeV/$c^{2}$ are rejected. The additional requirements  
$|M(\Lambda\pi^{-}/\bar{\Lambda}\pi^{+})-M(\Xi)|>0.008 $ GeV/$c^2$
and $|M(\Lambda\pi^{-}/\bar{\Lambda}\pi^{+})-M(\Sigma(1385))|>0.040 $ GeV/$c^2$ are applied to remove events with $\Xi$ and $\Sigma(1385)$ baryons in the final states, where $M(X)$ is the known mass of the $X$ particle from Particle Data Group (PDG)~\cite{2022pdg}.

\begin{figure}[htp]
 \centering
  \begin{overpic}[width=0.5\textwidth]{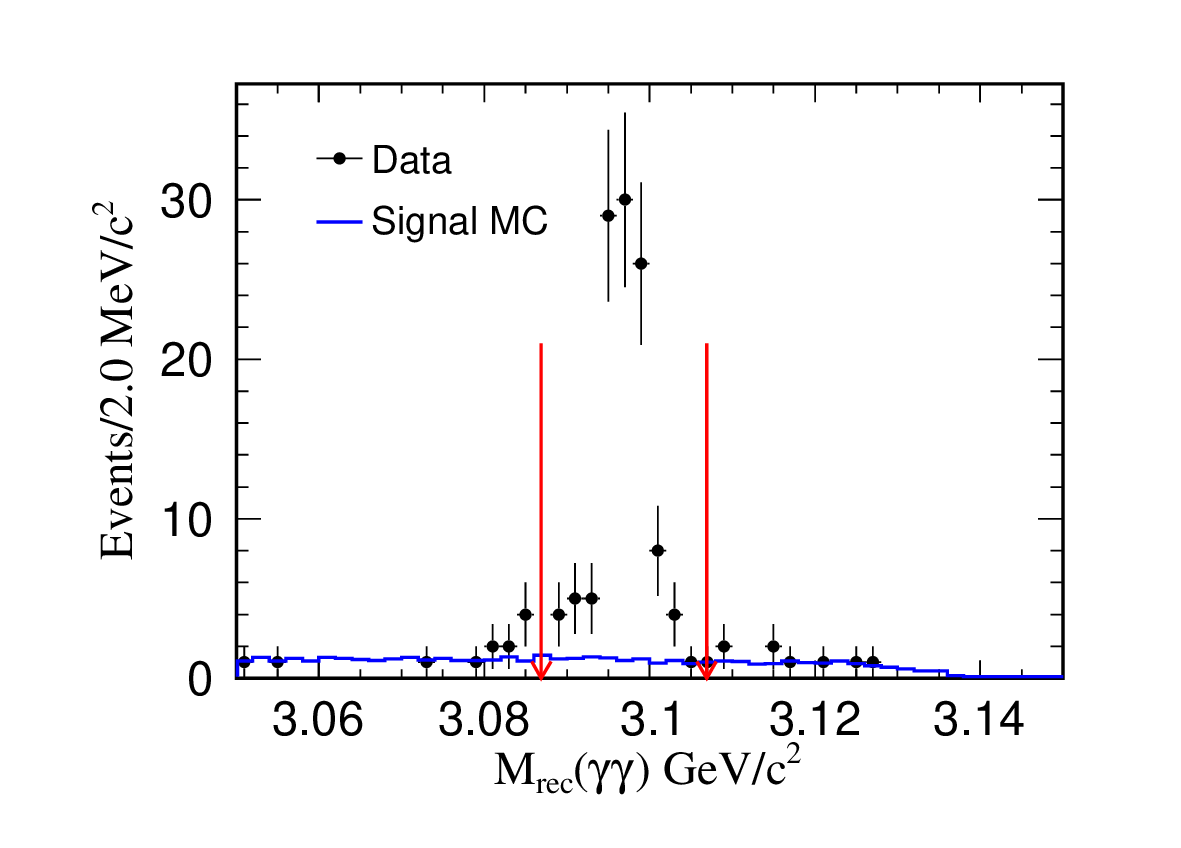}
  \end{overpic}
  \caption{Distributions of  $M_{\rm rec}(\gamma\gamma)$ for Mode II, where the solid arrows indicate the mass window within which $J/\psi$ decays are vetoed.  The signal MC is normalised to the measured branching fraction.
  }
  \label{fig:etapp_etarecoil}
\end{figure}

For  {\bf Mode II},  the $\Lambda(\Lamb)$ candidates are selected by requiring $1.108<M(p\pi^{-}/\bar{p}\pi^{+})<1.123 $ GeV/$c^{2}$. The background events from $\psi(3686)\to \eta\jpsi $  are removed by requiring $|M_{\rm rec}(\gamma \gamma)-M(\jpsi)|>10$ MeV/$c^{2}$, as shown in  Fig.~\ref{fig:etapp_etarecoil}, where  $M_{\rm rec}(\gamma \gamma)$ and $M(J/\psi)$ are the recoil mass against $\gamma \gamma$ and the mass of the $J/\psi$ meson from PDG, respectively.

\begin{figure}
\centering
\begin{overpic}[width=0.5\textwidth]{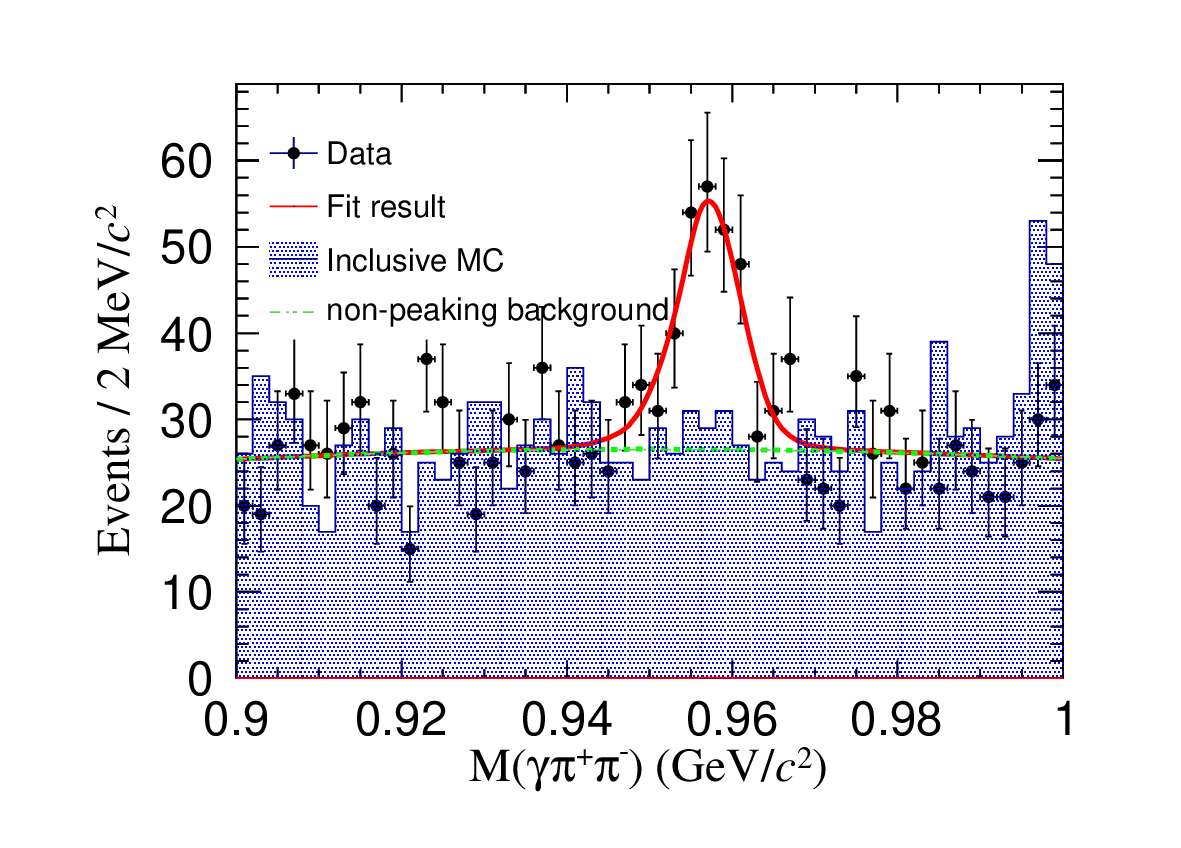}
  \put(70,55){$(a)$}
  \end{overpic}
 \begin{overpic}[width=0.5\textwidth]{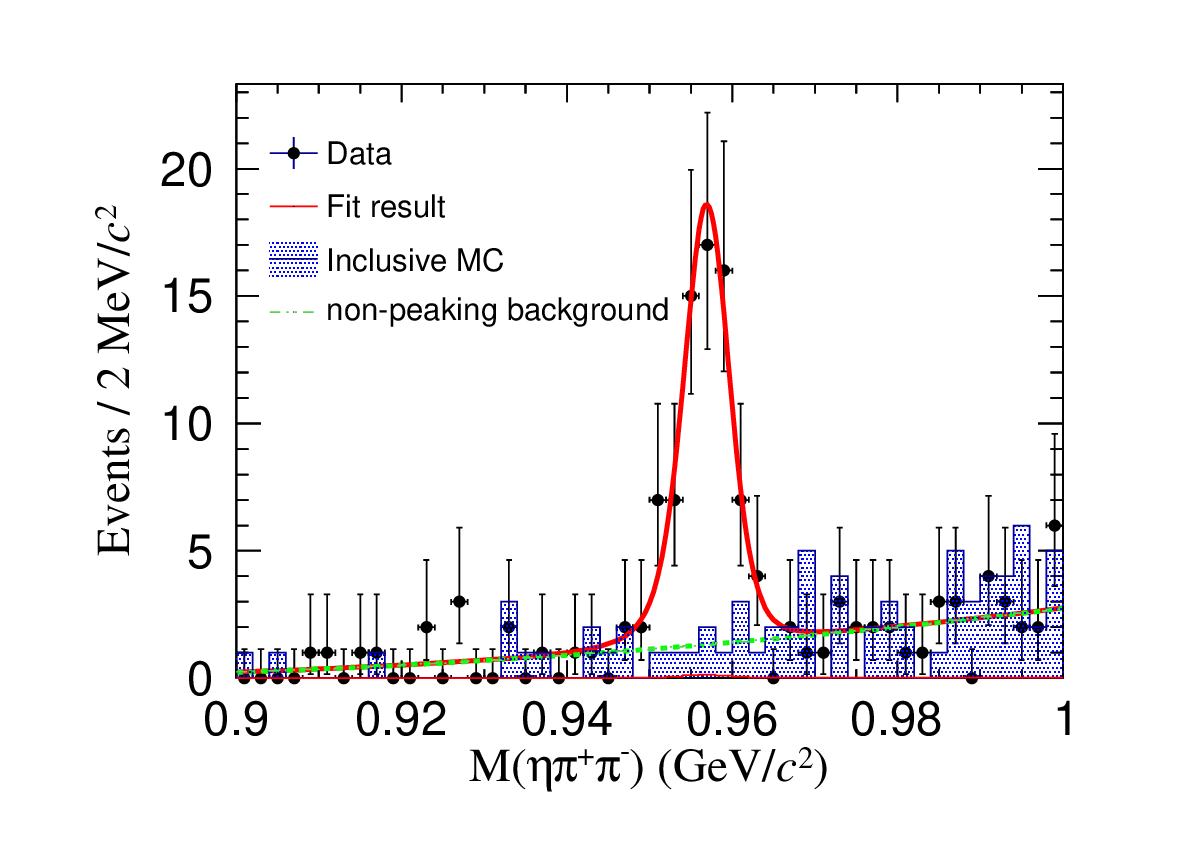}
  \put(70,55){$(b)$}
  \end{overpic}
  \caption{Distributions of $M(\gamma\pi^+\pi^-)$ (a) and $M(\eta \pi^+\pi^-)$ (b).}
 \label{fig:etapr}

\end{figure}

The $M(\gamma\pi^+\pi^-)$ and $M(\eta\pi^+\pi^-)$ distributions, after these selection requirements, are shown in Figs.~\ref{fig:etapr}(a) and~(b), where the  $\eta'$  peaks can be seen clearly. To verify that these peaks are not from background events,  the distribution of candidates  passing the same selection from  2.7 billion  inclusive $\psi(3686)$ events  are obtained and shown as shaded histograms in Figs.~\ref{fig:etapr}(a) and~(b);
the background events have an approximately flat distribution and do not peak in the $\eta^\prime$ mass region.

To estimate the number of
background events coming directly from $e^+e^-$ annihilation due to quantum electrodynamics processes, which we call QED background, 
the same analysis is performed on the data samples taken at the CM energy of 3.773 GeV, with an integrated luminosity of $(2916.94\pm0.18\pm29.17)$ pb$^{-1}$~\cite{ctn}. 
For {\bf Mode I}, several $\eta$ decays leak into $\eta'$ signal region in the $M(\gpp)$ distribution, and are classified as QED background.  The number of  background events is extracted by fitting the $M(\gpp)$ distribution, normalized to the $\psi(3686)$ data taking into account the integrated luminosity and energy dependent cross section of the QED processes~\cite{2012Ablikim4}. The number of QED background events is determined to be 
\begin{equation}
 N_{\rm bkg} = N_{\psi(3773)} \times 
 \frac{\mathcal{L}_{\psi(3686)}}{\mathcal{L}_{\psi(3773)}} \times \frac{s_{\psi(3773)}}{s_{\psi(3686)}},
\end{equation}
where $N_{\psi(3773)}$ is the number of background events obtained from data at the CM energy of 3.773 GeV; 
$\mathcal{L}_{\psi(3686)}$ and $\mathcal{L}_{\psi(3773)}$ is the integrated luminosity of $\psi(3686)$ and $\psi(3773)$ data samples, respectively, and $\sqrt{s_{\psi(3686)}}$ and $\sqrt{s_{\psi(3773)}}$ are the CM energies of $\psi(3686)$ and $\psi(3773)$ data samples, respectively. The normalized number of QED background events due to $e^{+}e^{-}$ annihilations is $8.1\pm4.8$ for {\bf Mode I}, which is subtracted in order to determine the signal yield. For {\bf Mode II}, only 2 events survive after the event selection, which is considered negligible.

\section{ Branching-fraction measurement}
The $\psi(3686)\to \LLE$ signal yields for {\bf Mode I} and {\bf Mode II} are obtained from an extended unbinned maximum likelihood fit to 
the $M(\gamma\pi^+\pi^-)$ and $M(\eta\pi^+\pi^-)$ distributions, respectively. The total probability density function 
consists of a signal and a non-peaking background contribution. The signal component is modeled with the MC signal shape convolved with a Gaussian function to account for  a possible difference in the mass resolution between data and MC simulation, and the non-peaking background is parameterized by a second-order Chebychev polynomial.
From the fit results, indicated by the red solid lines in Fig.~\ref{fig:etapr}, we obtain  $148\pm 24$ and $70\pm 10$ $\psi(3686)\rightarrow\Lambda\bar{\Lambda}\eta^\prime$ events with statistical significances of 6.8$\sigma$ and 10$\sigma$ for {\bf Mode I} and {\bf Mode II}, respectively.  The statistical significance is determined
by the change in the log-likelihood value and in the number of degrees of freedom in the fit with and without the $\eta^\prime$ signal.

\begin{figure}[htp]
 \centering
  \begin{overpic}[width=0.5\textwidth]{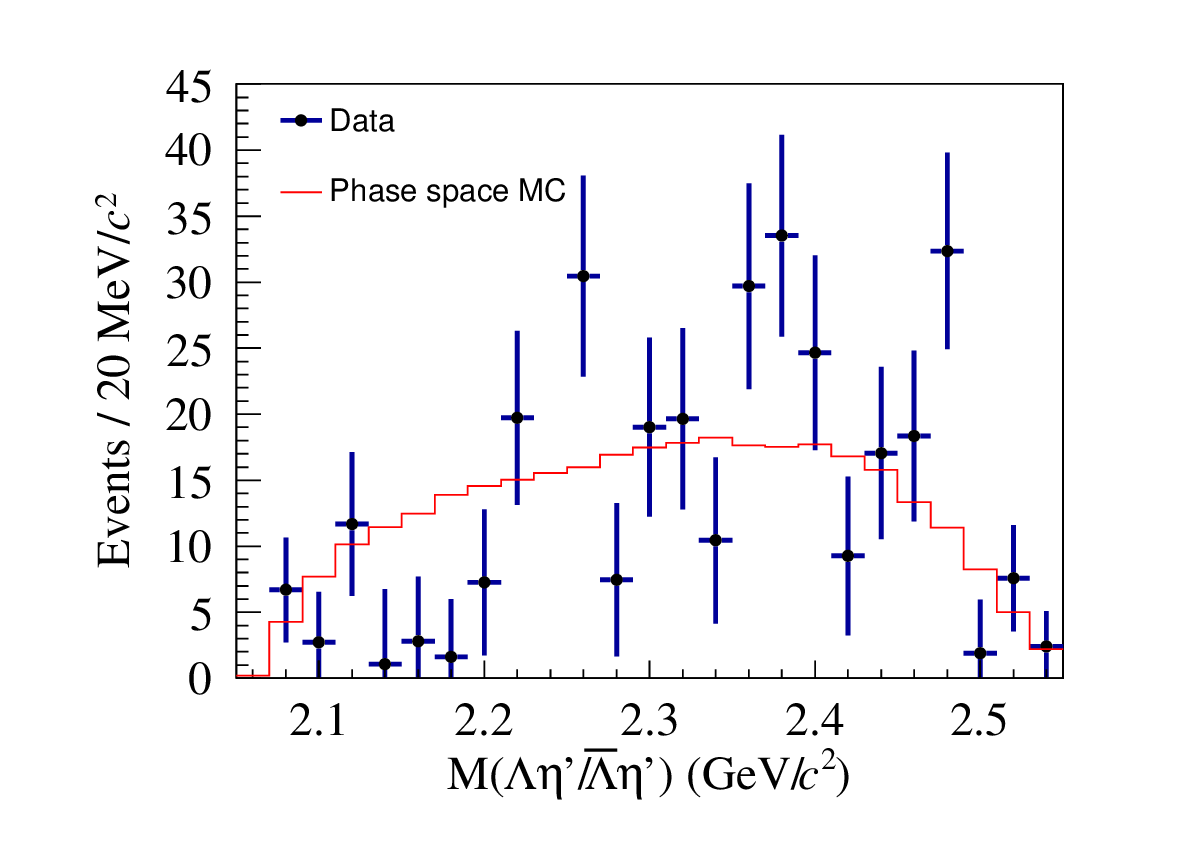}
    \put(30,48){$(a)$}
  \end{overpic}
  \begin{overpic}[width=0.5\textwidth]{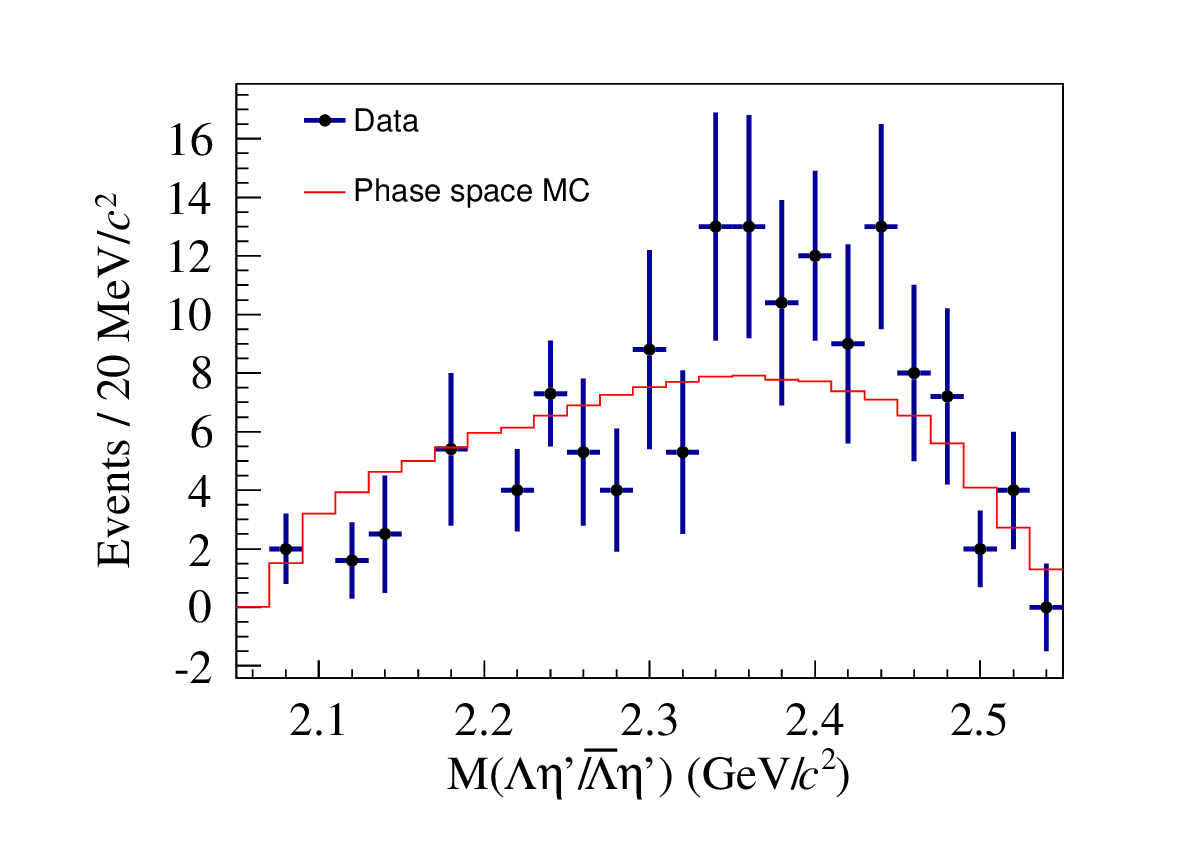}
      \put(30,48){$(b)$}
  \end{overpic}
  \caption{Distributions of $M_{\Lambda\eta'/\Lamb\eta'}$ for {\bf Mode I} (a) and {\bf Mode II} (b).}
  \label{fig:ext}
\end{figure}

\begin{figure}[htp]
 \centering
  \begin{overpic}[width=0.5\textwidth]{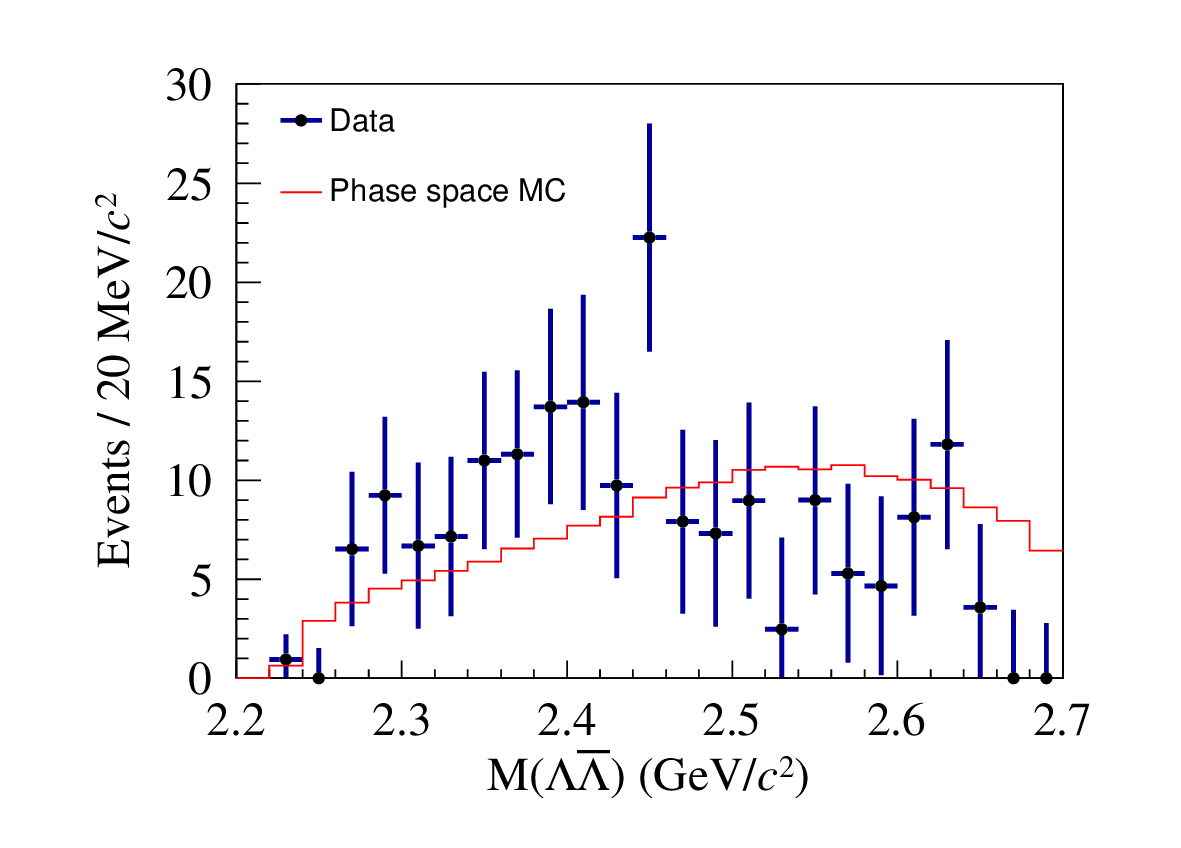}
    \put(70,55){$(a)$}
  \end{overpic}
  \begin{overpic}[width=0.5\textwidth]{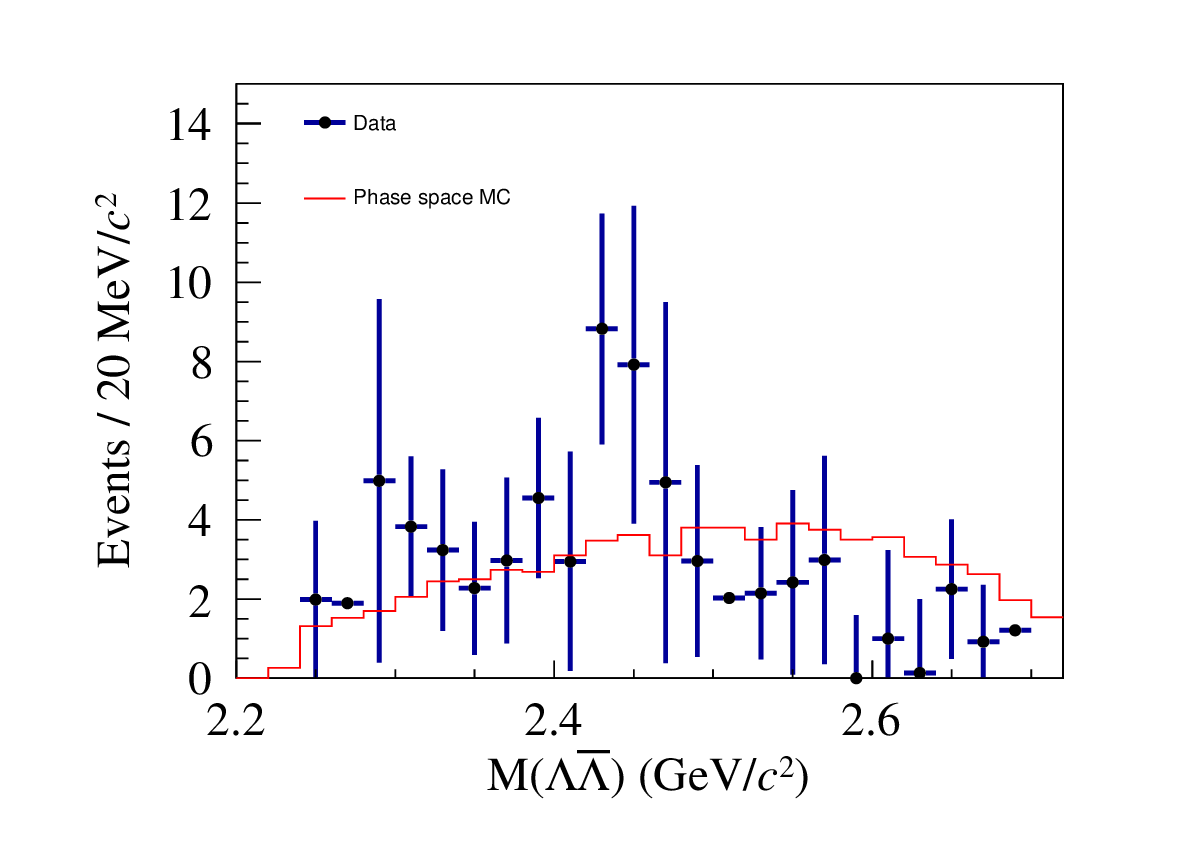}
    \put(70,55){$(b)$}
  \end{overpic}
  \caption{Distributions of $M_{\Lambda\Lamb}$ for {\bf Mode I} (a) and {\bf Mode II} (b).}
  \label{fig:ext-2lambda}
\end{figure}

To investigate for the presence of  possible intermediate structures, we study the background-subtracted   $\Lambda\eta^\prime (\bar{\Lambda}\eta^\prime)$ mass distributions. For each
bin, the number of signal events in data is extracted by fitting  the $M(\gamma\pi^+\pi^-)$ and $M(\eta\pi^+\pi^-)$ distributions, respectively, for {\bf Mode I} and {\bf Mode II}. The obtained results are shown in Fig.~\ref{fig:ext}, which display no evident structures. Furthermore, the background-subtracted $\Lambda \bar{\Lambda}$ mass distributions with the above similar procedure are shown in Fig.~\ref{fig:ext-2lambda}, where no structures are observed either.

The detection efficiencies are obtained from MC simulations
produced with a uniform phase-space distribution for the three-body decay $\psi(3686)\rightarrow \Lambda\bar{\Lambda}\eta^\prime$.
The detection efficiencies are  determined to be  $6.50\%$ and $4.58\%$ for {\bf Mode I}  and {\bf Mode II}, respectively.

The branching fraction of $\psi(3686)\rightarrow \Lambda\bar{\Lambda}\eta^\prime$ is calculated with
\begin{flalign}
\mathcal{B}(\psi(3686)\to\LLb\eta') 
 =\frac{N_{\rm obs}-N_{\rm bkg}}{\mathcal{N}_{\psi}  \mathcal{B}^2(\Lambda\to p\pi^{-})\mathcal{B}(\eta'\to X)  \varepsilon},
\end{flalign}
where $N_{\rm obs}$ is the number of observed signal candidates, $N_{\rm bkg}$ is the number of QED background events, $\mathcal{N}_{\psi}$  is  the  number of $\psi(3686)$ events~\cite{psinum}, 
$\varepsilon$ is the detection efficiency obtained from MC simulation,   
 $\mathcal{B}(\Lambda\to p\pi^{-})$ is the branching fraction of $\Lambda\to p\pi^{-}$ and
  $\mathcal{B}(\eta'\to X) $ represents the branching fraction of $\eta'\to\gamma\pi^{+}\pi^{-}$ or the product branching fraction of $\eta'\to\eta\pi^{+}\pi^{-}$ and $\eta\to\gamma\gamma$. The intermediate branching fractions are taken from the PDG~\cite{2022pdg}.

\begin{table}[htp]
\caption{The branching fraction results and the values used in the branching fraction calculation for each decay mode, where the first uncertainty is statistical
 and the second is systematic.}
\centering 
\begin{tabular}{ccccc}\\\hline\hline
 Mode & $N_{\rm obs}$& $N_{\rm bkg}$ & $\varepsilon$ $(\%)$ &$\mathcal{B}$ ($\times 10^{-6}$)\\\hline
I&$148\pm24$&$8.1\pm 4.8$ &6.50& $6.59\pm1.15\pm0.53$\\
 II &$70\pm10$ & &4.58& $8.25\pm1.18\pm0.67$\\
\hline
 Combined & & & & $7.34\pm0.94\pm0.43$\\\hline
 \hline
\end{tabular}
\label{tab:com}
\end{table}

The obtained values of the branching fractions of $\psi(3686)\rightarrow \Lambda\bar{\Lambda}\eta^\prime$
are summarized in Table~\ref{tab:com}. The results
from the two $\eta'$ decay modes are consistent with each
other within their uncertainties, thus the two measurements 
are combined, taking accounted of  the correlated uncertainties (see section~\ref{syst} for details); the mean value and the uncertainty are
calculated following the procedure of Ref.~\cite{Agostini-1994},
\begin{equation}
    \bar{x}\pm \sigma(\bar{x}) = \frac{\Sigma_{j}(x_{j}\cdot \Sigma_{i} w_{ij})}{\Sigma_{i}\Sigma_{j}w_{ij}}
    \pm \sqrt{\frac{1}{\Sigma_{i}\Sigma_{j}w_{ij}}},
\end{equation}
where $i$ and $j$ are summed over all decay modes, $w_{ij}$ is
the element of the weight matrix $W = V^{-1}_{x}$
, and $V_{x}$ is
the covariance error matrix calculated according to the
statistical uncertainties  and the systematic uncertainties. When combining the results of the two decay modes, the error matrix is calculated as 
 \begin{equation}
 \label{xxx}
     V =
     \begin{pmatrix}
    \sigma_{1}^{2}+\epsilon_{f}^{2} x_{1}^{2}   & \epsilon_{f}^{2}x_{1}x_{2}\\
    \epsilon_{f}^{2}x_{1}x_{2} &      \sigma_{2}^{2}+\epsilon_{f}^{2}x_{2}^{2} 
     \end{pmatrix},
\end{equation}
where $\sigma_{i}$
is the independent absolute uncertainty (the
statistical uncertainty and all independent systematical
uncertainties are added in quadrature) in the measurement
mode $i$, and $\epsilon_{f}$ is the common relative systematic uncertainty between the two measurements
(all the common systematic uncertainties are added in quadrature; the entries in Table \ref{tab:unt1} marked with  $^{'*'}$ are the uncertainties in common with the two $\eta^{\prime}$ decays while the other uncertainties are different); $x_{i}$
is the measured value for mode $i$.
 The combined branching fraction of $\psi(3686)\rightarrow \Lambda\bar{\Lambda}\eta^\prime$ is found to be $(7.34\pm0.94(stat.)\pm0.43(sys.))\times 10^{-6}$.

 \section{Systematic uncertainties}
 \label{syst}
The sources of systematic uncertainties and their corresponding contributions to the measurements of the branching fractions are summarized in Table \ref{tab:unt1}. Assuming that all sources are independent, the total systematic uncertainties are obtained by adding the individual contributions in quadrature
and are determined
to be
$8.0\%$ for {\bf Mode I} and $8.1\%$ for {\bf Mode II}.

{\bf Number of $\psi(3686)$ events:} 
The total number of $\psi(3686)$ events, $(27.12\pm0.14)\times 10^{8}$, is 
determined by measuring the yield of inclusive hadronic events~\cite{psinum}, and its uncertainty is estimated to be 0.5\%.

{\bf Tracking efficiency:}
The uncertainty due to differences in the tracking efficiency between data and MC is 1.0\%
for each charged track coming from a primary vertex, according to a study of $J/\psi\to \rho\pi$ and $J/\psi \to p\bar{p}\pi^{+}\pi^{-}$ events~\cite{2012Ablikim6}.  Therefore, the uncertainty from this source associated with the $\eta'$ reconstruction is 2.0\%.   The corresponding uncertainty associated with the $\Lambda$ and $\bar\Lambda$ reconstruction is determined separately, and discussed below.

{\bf PID efficiency: }
The PID efficiency has been investigated using a control sample of  $J/\psi \to p\bar{p}\pi^{+}\pi^{-}$ decays. The uncertainty is assigned to be 1.0\% 
per charged track. In this analysis, all charged tracks must satisfy
PID requirements, so the total systematic uncertainty from this source is assigned to be 6.0\%.\par
{\bf Photon detection efficiency: }
The efficiency of the photon reconstruction is studied with a 
control sample $\psi(3686)\to \pi^{+}\pi^{-}J/\psi,J/\psi \to \rho^{0} \pi^{0}$ events~\cite{2011Ablikim8}. The systematic uncertainty in the photon selection is assigned to be 1.0\% per photon.\par

{ \bf $\Lambda$ and $\bar{\Lambda}$ reconstruction efficiency:} The momentum-dependent $\Lambda(\bar{\Lambda})$ reconstruction efficiency has been studied by calculating the overall tracking efficiency in a given $\cos\theta$ range using a control sample of $J/\psi \to p K^- \bar{\Lambda} + c.c.$ decays.
The difference between data and MC simulation, 0.70\%, is taken as the systematic uncertainty. 

{ \bf Kinematic fit: }
The systematic uncertainty due to the kinematic fit is estimated by correcting the
helix parameters
of charged tracks according to the method described in Ref.~\cite{2013Ablikim11}. We take the efficiency from the track-parameter-corrected MC sample as the baseline value, and assign half of the difference in the signal efficiencies before and after the correction as the associated systematic uncertainty, which is  0.6\%  for both {\bf Mode I} and {\bf Mode II}. \par

{\bf Mass window:}
 The uncertainty due to the mass windows used to select signal events or veto backgrounds originates from the differences in the mass resolutions between data and MC simulation. The
analysis is repeated with larger and smaller values of the mass window.
 The maximum relative change in the measured  branching fraction is not significant after considering the correlations between the signal yields,
hence this
uncertainty is considerred to be negligible.

{\bf Intermediate decays: }
The systematic uncertainties associated with the knowledge of the branching fractions of the intermediate decays, including $\Lambda\rightarrow p\pi^-$, $\eta^\prime\rightarrow\gamma\pi^+\pi^-$, $\eta^\prime\rightarrow \pi^+\pi^-\eta$ and $\eta\rightarrow\gamma\gamma$,  are taken from the uncertainties listed in the PDG~\cite{2022pdg}.\par

{ \bf Continuum background: }
The systematic uncertainty associated with the level of continuum background is determined by re-determining the branching fraction after varying the continuum background yield by 
$\pm 1\sigma$ of its statistical uncertainty and assigning the change in branching fraction as the systematic uncertainty.\par

{\bf Signal shape and background shape: } 
 To estimate the uncertainty due to the choice of signal shape, the  MC-simulated shape convolved with a  Gaussian function is replaced by a simple MC shape and the resulting differences in the branching fraction is assigned as the systematic uncertainty.  In the case of the background shape, the second-order Chebychev polynomial used for the baseline result is replaced by a first-order or third-order Chebychev polynomial.  The largest change of branching fraction is taken as the systematic uncertainty.
 
{\bf Fit range: } The uncertainty associated with this source is estimated by varying the fit range as 
 (0.89,1.01) GeV/$c^{2}$, (0.89,0.99) GeV/$c^{2}$, (0.91,1.01) GeV/$c^{2}$ and (0.91,0.99) GeV/$c^{2}$, and the maximum resulting difference is assigned as the systematic uncertainty.

 \begin{table}[!htbp]

 \caption{\small Relative systematic uncertainties in the branching-fraction measurement (in unit of $\%$).
The sources marked '*' are in common for the two $\eta^\prime$ decay modes. }
 \centering 
\begin{tabular}{lcc}
\hline \hline

 Sources & Mode I & Mode II\\\hline

 Number of $\psi(3686)$ events* & 0.5&0.5\\
 MDC tracking* &2.0&2.0\\
 PID efficiency* &6.0&6.0\\
 Photon detection efficiency &1.0&2.0\\
 $\Lambda(\bar{\Lambda})$ reconstruction* &0.7&0.7\\
 Kinematic fit & 0.6&0.6\\
 \hline
$\Lambda(\bar{\Lambda})$ intermediate decays &  &  \\
$\Lambda\to p\pi^{-}$* & \multirow{2}*{1.6}& \multirow{2}*{1.6} \\
 $\bar{\Lambda}\to \bar{p}\pi^{+}$*  \\\hline
$\eta'$ intermediate decays &  &  \\
 $\eta'\to \gamma\pi^{+}\pi^{-}$ & 1.4&-\\
 $\eta'\to \eta\pi^{+}\pi^{-}$ &-&1.2\\
 $\eta\to \gamma \gamma $ &-& 0.5 \\
 \hline
 Continuum background & 3.3&-\\
 \hline
Mass spectrum fitting &  &  \\
 Signal shape & 0.7 &0.7\\
 Background shape & 2.0 &1.2\\
 Fit range & 1.4 &3.7\\
 \hline
 Total  &8.0&8.1\\
 \hline \hline
\end{tabular}
\label{tab:unt1}
\end{table}

\section{Summary}

The decay $\psi(3686) \to \Lambda \bar{\Lambda}\eta'$ with the subsequent decay modes $\eta^\prime\rightarrow\gamma\pi^+\pi^-$ ({\bf Mode I}) and  $\eta^\prime\rightarrow\eta\pi^+\pi^-$ ({\bf Mode II})
is observed for the first time, using a data sample of 
$(27.12\pm 0.14) \times 10^{8}$ 
$\psi(3686)$ events. The corresponding branching fractions are measured be $(6.59\pm1.15(stat.)\pm0.53(sys.)) \times 10^{-6} $ and $(8.25\pm1.18(stat.)\pm0.67(sys.)) \times 10^{-6}$, respectively. The combined branching fraction is  $B(\psi(3686) \to \Lambda \bar{\Lambda}\eta')=(7.34\pm0.94(stat.)\pm0.43(sys.)) \times 10^{-6}$.  

We have also searched for possible excited $\Lambda$ states  but no evident structure is observed.
In the future, a super $\tau$ charm factory will collect more data and searches for excited hyperon states will become possible~\cite{Achasov2023gey}.

\section{acknowledgments}

The BESIII Collaboration thanks the staff of BEPCII and the IHEP computing center for their strong support. This work is supported in part by National Key R\&D Program of China under Contracts Nos. 2020YFA0406300, 2020YFA0406400; National Natural Science Foundation of China (NSFC) under Contracts Nos. 11635010, 11735014, 11835012, 11935015, 11935016, 11935018, 11961141012, 12022510, 12025502, 12035009, 12035013, 12061131003, 12075250, 12192260, 12192261, 12192262, 12192263, 12192264, 12192265, 12221005, 12225509, 12235017; the Chinese Academy of Sciences (CAS) Large-Scale Scientific Facility Program; the CAS Center for Excellence in Particle Physics (CCEPP); Joint Large-Scale Scientific Facility Funds of the NSFC and CAS under Contract No. U1832207; CAS Key Research Program of Frontier Sciences under Contracts Nos. QYZDJ-SSW-SLH003, QYZDJ-SSW-SLH040; 100 Talents Program of CAS; The Institute of Nuclear and Particle Physics (INPAC) and Shanghai Key Laboratory for Particle Physics and Cosmology; European Union's Horizon 2020 research and innovation programme under Marie Sklodowska-Curie grant agreement under Contract No. 894790; German Research Foundation DFG under Contracts Nos. 455635585, Collaborative Research Center CRC 1044, FOR5327, GRK 2149; Istituto Nazionale di Fisica Nucleare, Italy; Ministry of Development of Turkey under Contract No. DPT2006K-120470; National Research Foundation of Korea under Contract No. NRF-2022R1A2C1092335; National Science and Technology fund of Mongolia; National Science Research and Innovation Fund (NSRF) via the Program Management Unit for Human Resources \& Institutional Development, Research and Innovation of Thailand under Contract No. B16F640076; Polish National Science Centre under Contract No. 2019/35/O/ST2/02907; The Swedish Research Council; U. S. Department of Energy under Contract No. DE-FG02-05ER41374.



\newpage
\begin{thebibliography}{99}

\bibitem{Kwong1987mj}
W. Kwong, J. L. Rosner and C. Quigg,
\href{https://inspirehep.net/files/92cd561c57c96570c626e50005a0e44e}{Ann. Rev. Nucl. Part. Sci. \textbf{37}, 325 (1987).}


\bibitem{Eichten2007qx}
E. Eichten, S. Godfrey, H. Mahlke and J. L. Rosner,
\href{https://journals.aps.org/rmp/pdf/10.1103/RevModPhys.80.1161}{Rev. Mod. Phys. \textbf{80}, 1161 (2008).}


\bibitem{HP}
K. Zhu, X. H. Mo and C. Z. Yuan,
\href{https://www.worldscientific.com/doi/abs/10.1142/S0217751X15501481}{Int. J. Mod. Phys. A 30, 1550148 (2015).}


\bibitem{2022ws}
M. Ablikim \textit{et al}. [BESIII Collaboration],
\href{https://journals.aps.org/prd/abstract/10.1103/PhysRevD.106.072006}{Phys. Rev. \textbf{D 106}, 072006 (2022).}

\bibitem{2013ablikim}
M. Ablikim \textit{et al}. [BESIII Collaboration],
\href{https://journals.aps.org/prd/abstract/10.1103/PhysRevD.87.052007}{Phys. Rev.\textbf{D 87}, 052007 (2013).}


\bibitem{ex}
A. V. Sarantsev, M. Matveev, V. A. Nikonov,  
A. V. Anisovich, U. Thoma and E. Klempt, \href{https://doi.org/10.1140/epja/i2019-12880-5}{Eur. Phys. J.\textbf{ A 55}, 180 (2019).}


\bibitem{Hex-2020}
Volker Crede,   
\href{https://pubs.aip.org/aip/acp/article/2249/1/020003/1002149/The-excited-baryon-spectrum-What-have-we-learned}{AIP Conf. Proc. \textbf{2249}, 020003 (2020)}

\bibitem{psinum}
 M.~Ablikim {\it et al.} [BESIII Collaboration],
\href{https://doi.org/10.1088/1674-1137/42/2/023001}{Chin. Phys. C {\bf 42}, 023001 (2018).}
With the same approach as for $\psi(3686)$ events taken in 2009, the preliminary number of
$\psi(3686)$ events taken in 2009, 2012 and 2021 is determined to be $27.12\times 10^{8}$ with an uncertainty of 0.5\%.

\bibitem{2022pdg}
R. L. Workman {\it et al}. [Particle Data Group],
\href{https://doi.org/10.1093/ptep/ptac097}{Prog. Theor. Exp. Phys. \textbf{2022}, 083C01 (2022).}

\bibitem{Ablikim:2009aa}
  M.~Ablikim {\it et al.} [BESIII Collaboration],
  \href{https://arxiv.org/ftp/arxiv/papers/0911/0911.4960.pdf}{Nucl.\ Instr.\ Meth.\ Phys.\ Res.\ Sect.\ A {\bf 614}, 345 (2010).}

\bibitem{Yu:IPAC2016-TUYA01}
   C.~H.~Yu {\it et al.},
  \href{http://jacow.org/ipac2016/papers/tuya01.pdf}{
  Proceedings of IPAC2016, Busan, Korea, 2016.}
 
  \bibitem{Ablikim:2019hff}
  M.~Ablikim {\it et al.} [BESIII Collaboration],
  \href{https://iopscience.iop.org/article/10.1088/1674-1137/44/4/040001}{Chin. Phys. C {\bf 44}, 040001 (2020).}

\bibitem{EventFilter}
  J.~W.~Zhang, L.~H.~Wu, S.~S.~Sun {\it et al.},
 \href{https://link.springer.com/article/10.1007/s41605-022-00331-7}{Radiat. Detect. Technol. Methods {\bf 6}, 289–293 (2022).}


\bibitem{tof2017lix} X. Li \textit{et al.}, 
\href{https://link.springer.com/article/10.1007/s41605-017-0014-2}{Radiat. Detect. Technol. Methods \textbf{1}, 13  (2017).}
\bibitem{tof2017caop} Y. X. Guo \textit{et al.}, 
\href{https://link.springer.com/article/10.1007/s41605-017-0012-4}{Radiat. Detect. Technol. Methods \textbf{1}, 15 
(2017).}
\bibitem{tof2020caop} P. Cao \textit{et al.}, 
\href{https://www.sciencedirect.com/science/article/pii/S0168900219314068?via%3Dihub}{
Nucl. Instrum. Methods Phys. Res., Sect. \textbf{ A
953}, 163053 (2020).}


\bibitem{2003Agostinelli}
S. Agostinelli  \textit{et al.} [GEANT4 collaboration],
\href{https://www.sciencedirect.com/science/article/abs/pii/S0168900203013688?via\%3Dihub}{Nucl. Instrum. Meth.  \textbf{A 506}, 250 (2003).}

\bibitem{kkmc2000} 
S. Jadach, and B. F. L. Ward and Z. Was, 
\href{https://www.sciencedirect.com/science/article/abs/pii/S0010465500000485?via\%3Dihub}{Comput. Phys. Commun. \textbf{ 130} 260 (2000).}

\bibitem{kkmc2001} 
S. Jadach, B. F. L. Ward and Z. Was,
\href{https://arxiv.org/pdf/hep-ph/0006359.pdf}{Phys. Rev.  \textbf{D 63}, 113009 (2001).}


\bibitem{2001Lange}
D. J. Lange,
\href{https://www.sciencedirect.com/science/article/pii/S0168900201000894?via\%3Dihub}{Nucl. Instrum. Meth. \textbf{A 462}, 152 (2001).}

\bibitem{2008ping}
R. G. Ping,
\href{https://iopscience.iop.org/article/10.1088/1674-1137/32/8/001}{ Chin. Phys. \textbf{C 32}, 599 (2008).}

 
\bibitem{2000chen}
J. C. Chen, G. S. Huang, X. R. Qi, D. H. Zhang and Y. S. Zhu,
\href{https://journals.aps.org/prd/abstract/10.1103/PhysRevD.62.034003}{Phys. Rev.  \textbf{D 62}, 034003 (2000).}
\bibitem{lund2014}
 R. L. Yang, R. G. Ping, and H. Chen,
\href{https://iopscience.iop.org/article/10.1088/0256-307X/31/6/061301}{Chin. Phys. Lett. \textbf{ 31},
061301 (2014).}

\bibitem{mode1-gammapipi}
J. Wess and B. Zumino, \href{https://doi.org/10.1016/0370-2693(71)90582-X}{Phys. Lett. B \textbf{37}, 95 (1971).}

\bibitem{mode2-gammapipi}
E. Witten, 
\href{https://doi.org/10.1016/0550-3213(83)90063-9}{Nucl. Phys. B \textbf{ 223}, 422 (1983).}

\bibitem{bes3-gammapipi}
M. Ablikim  {\it et al.} [BESIII Collaboration], \href{https://journals.aps.org/prl/pdf/10.1103/PhysRevLett.120.242003}{Phys. Rev. Lett. textbf{ 120}, 242003 (2018).}

\bibitem{mode1-etaptoetapipi}
M. Ablikim {\it et al.} [BESIII Collaboration], 
\href{https://journals.aps.org/prd/pdf/10.1103/PhysRevD.97.012003}{Phys. Rev. D \textbf{ 97}, 012003 (2018).}

\bibitem{ctn}
M. Ablikim {\it et al.} [BESIII Collaboration], 
\href{https://iopscience.iop.org/article/10.1088/1674-1137/37/12/123001}{Chin. Phys. \textbf{C 37}, 123001 (2013).}
\bibitem{ctn2}
M. Ablikim {\it et al.} [BESIII Collaboration], 
\href{ https://inspirehep.net/files/90f19f568cfbc1b1cfe154e1a3737876}{Phys. Lett. B \textbf{753}, 629 (2016). }
 

\bibitem{2012Ablikim4}
M. Ablikim {\it et al.}  [BESIII Collaboration],
\href{https://iopscience.iop.org/article/10.1088/1674-1137/36/10/001}{Chin. Phys.  \textbf{C 36}, 915 (2012).}


\bibitem{Agostini-1994}
G. D’Agostini, \href{https://doi.org/10.1016/0168-9002(94)90719-6}{Nucl. Instrum. Methods Phys. Res., Sect. \textbf{ A
346}, 306 (1994).}


\bibitem{2012Ablikim6}
M. Ablikim {\it et al.}  [BESIII Collaboration],
\href{https://journals.aps.org/prd/abstract/10.1103/PhysRevD.85.092012}{ Phys. Rev. \textbf{D 85}, 092012 (2012). }
 
 
\bibitem{2011Ablikim8}
M. Ablikim {\it et al.}  [BESIII Collaboration],
\href{https://journals.aps.org/prd/abstract/10.1103/PhysRevD.83.112005}{Phys. Rev. {\bf D 83}, 112005 (2011).}


\bibitem{2013Ablikim11}
M. Ablikim \textit{et al.} [BESIII Collaboration],
\href{https://journals.aps.org/prd/abstract/10.1103/PhysRevD.87.012002}{Phys. Rev. {\bf D 87}, 012002 (2013).}


\bibitem{Achasov2023gey}
M. Achasov \textit{et al.},  
\href{https://arxiv.org/abs/2303.15790}{arXiv:2303.15790.}


\end{thebibliography}
\end{document}